\documentclass[12pt]{article}
\usepackage{graphicx,psfrag,epsf}
\usepackage{enumerate}
\usepackage{natbib}
\usepackage{url} 
\usepackage{amsmath,bbm,amssymb}
\usepackage{amsthm}
\usepackage{comment}
\usepackage{color}

\newcommand{\blind}{0}

\newtheorem{theorem}{Theorem}

\newtheorem{proposition}{Proposition}

\begin{document}

\def\spacingset#1{\renewcommand{\baselinestretch}%
{#1}\small\normalsize} \spacingset{1}


\if0\blind
{
  \title{\bf Rotated sphere packing designs}
  \author{Xu He\thanks{
    The authors are grateful to the referees, associate editor, Jeff C. F. Wu, V. Roshan Joseph and Peter Z. G. Qian for their valuable comments. 
He's work is supported by NSFC 11501550.}\hspace{.2cm}\\
    Academy of Mathematics and System Sciences, \\Chinese Academy of Sciences}
  \maketitle
} \fi

\if1\blind
{
  \bigskip
  \bigskip
  \bigskip
  \begin{center}
    {\LARGE\bf Rotated sphere packing designs}
\end{center}
  \medskip
} \fi

\bigskip
\begin{abstract}
We propose a new class of space-filling designs called rotated sphere packing designs for computer experiments.
The approach starts from the asymptotically optimal positioning of identical balls that covers the unit cube. 
Properly scaled, rotated, translated and extracted, such designs are excellent in maximin distance criterion, low in discrepancy, good in projective uniformity and thus useful in both prediction and numerical integration purposes. 
We provide a fast algorithm to construct such designs for any numbers of dimensions and points with R codes available online. 
Theoretical and numerical results are also provided. 
\end{abstract}

\noindent%
{\it Keywords:}  Experimental Design, Lattice, Low discrepancy, Maximin distance design, Minimax distance design 
\vfill

\newpage
\spacingset{1.45} 
\section{Introduction}
\label{sec:intro}

Space-filling designs are popular for computer experiments~\citep{Sacks:1989, Santner:2003}. 
It is commonly believed that design points should be evenly spread in the experimental space. 
In this paper, we consider the problem of choosing $n$ input vectors in the region $[0,1]^p$ for arbitrary $p$ and $n$. 
Our goal is to develop a new class of space-filling design that is suitable for both prediction and numerical integration purposes. 

\begin{figure}
\begin{center}
\includegraphics[width=14cm]{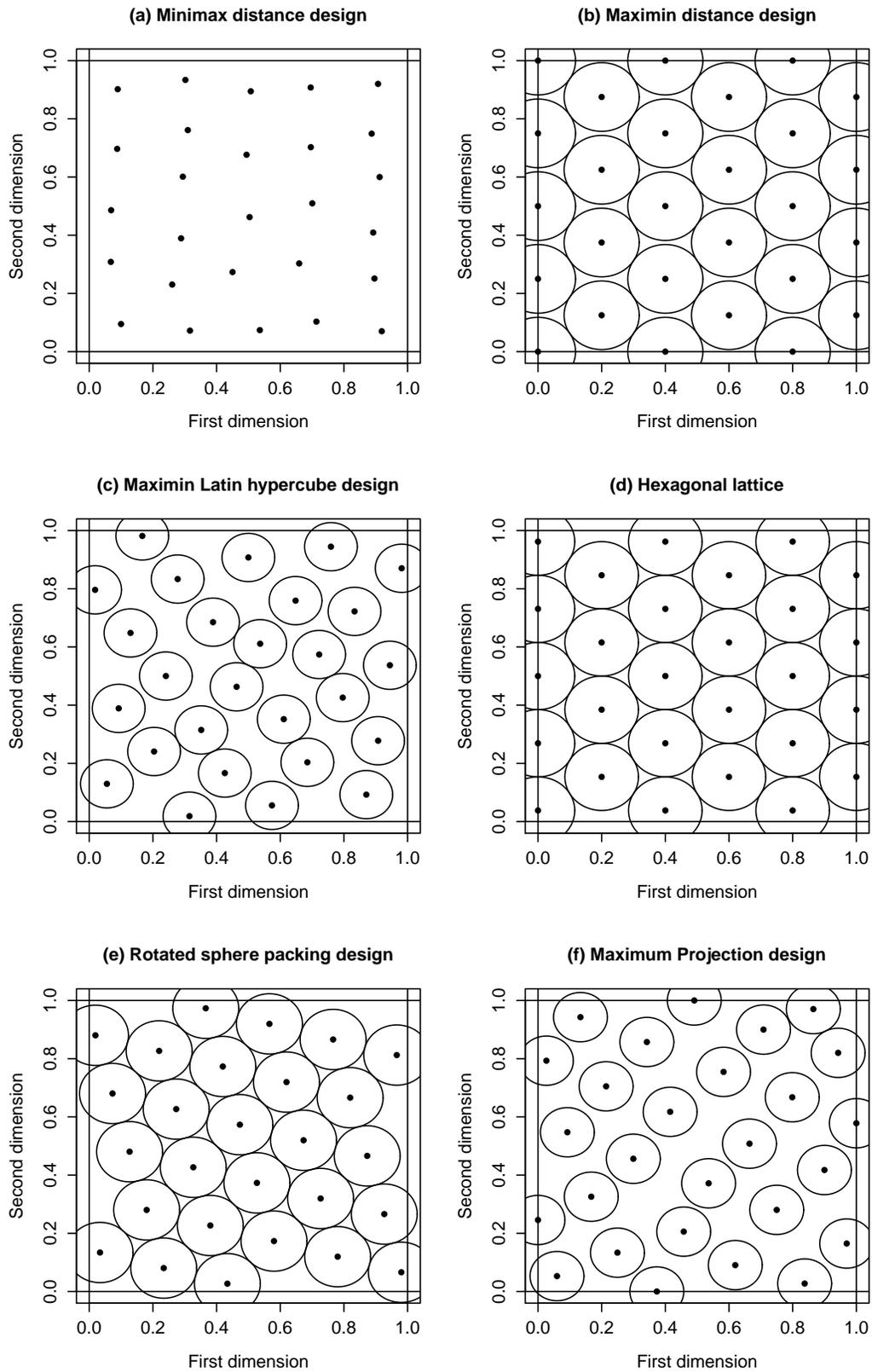}
\caption{Six two-dimensional space-filling designs with distance properties, $n=27$. \label{fig:intro}}
\end{center}
\end{figure}

Our idea comes from the minimax distance designs with the $L_2$ distance~\citep{Johnson:1990}. 
Let $\mathbf{D}$ denote a set of $n$ inputs in $[0,1]^p$. 
Then $\mathbf{D}$ is said to be a minimax distance design if it minimizes the worst predictive distance, 
\begin{equation}\label{eqn:MaxDist}
 \sup_{\mathbf{z}\in [0,1]^p}\left\{ \min_{\mathbf{x}\in \mathbf{D}}(\|\mathbf{z}-\mathbf{x}\|_2) \right\}, 
\end{equation}
where $\|(z_1,\ldots,z_p)\|_2 = (\sum_{k=1}^p z_k^2)^{1/2}$ is the $L_2$ norm. 
When a Gaussian process model or other nonlinear functions are used as a surrogate model for computer experiments, 
the prediction error of the output at an input $\mathbf{z}$ is closely related to its distance to nearest design point, $\min_{\mathbf{x}\in \mathbf{D}}(\|\mathbf{z}-\mathbf{x}\|_2)$. 
Therefore, we can reduce and control prediction error by minimizing the worst predictive distance. 
Although minimax distance designs are intuitively reasonable and useful, 
the challenge is how to construct such designs. 
Existing work includes~\citet{John:1995} for minimax distance two level factorial designs, 
\citet{Dam:2008} for two-dimensional minimax distance Latin hypercube designs and  
\citet{Tan:2013} for minimax distance designs in finite design spaces. 
Recently, \citet{Mak:2016} proposed to generate minimax distance designs by clustering.
Figure~\ref{fig:intro}(a) displays a minimax distance design with $p=2$ and $n=27$.

A related design criterion is the maximin distance~\citep{Johnson:1990}.
A design $\mathbf{D}$ is said to be a maximin distance design if it maximizes the minimum pairwise distance, 
\begin{equation}\label{eqn:MinDist}
 \min_{\mathbf{x}_1,\mathbf{x}_2\in \mathbf{D},\mathbf{x}_1\neq \mathbf{x}_2}(\|\mathbf{x}_1-\mathbf{x}_2\|_2) .
\end{equation}
Maximin distance designs are much more popular than minimax distance designs because they are easier to construct. 
It is also shown in~\citet{Johnson:1990} that maximin distance designs are usually good in the minimax sense. 
The website \url{http://www.packomania.com/} gives numerous best known maximin distance designs in two and three dimensions. 
Figure~\ref{fig:intro}(b) displays a maximin distance design with $p=2$ and $n=27$ from the website. 
Other types of designs with distance criteria include maximum entropy designs~\citep{Shewry:1987}, minimum energy designs~\citep{Joseph:2015}, among others. 

Designs with distance properties are usually good in prediction accuracy. 
However, if it turns out that only a subset of input variables are relevant in predicting the response, then prediction error is related to uniformity of the projected designs. 
Most maximin distance designs have coincident entries in their projections and are therefore poor in projective uniformity. 
Besides, they are also unsuitable for the integration purpose.  
In numerical integration, the mean output is usually estimated by the average of outputs. 
It is observed in Figure~\ref{fig:intro}(b) that many design points are located on the boundary of the design space. 
Let the Voronoi cell of a point $\mathbf{x}_i\in \mathbf{D}$ be the region $\text{Vor}(\mathbf{x}_i) = \{\mathbf{z} : \|\mathbf{z}-\mathbf{x}_i\|_2 = \min_{\mathbf{x}\in \mathbf{D}} \|\mathbf{z}-\mathbf{x}\|_2 \}$. 
Then clearly the inner points have much larger Voronoi cells than the boundary points. 
As a result, the average of outputs is a biased estimator for the mean response. 

To overcome these deficiencies and for easier optimization, the maximin distance criterion is usually used in combination with the Latin hypercube constraint. 
There is a vast literature on maximin distance Latin hypercube designs including: \citet{Morris:1995}, \citet{Jin:2005}, \citet{Liefvendahl:2006}, \citet{Dam:2007}, \citet{Grosso:2009}, among others. 
Figure~\ref{fig:intro}(c) displays a maximin distance Latin hypercube design with $p=2$ and $n=27$, generated from a simulated annealing algorithm~\citep{Morris:1995}. 
While losing slightly in minimum pairwise distance, maximin distance Latin hypercube designs hold excellent projective uniformity and are suitable for the integration purpose. 

The minimax and maximin distance designs are closely related to the mathematical problem of placing identical balls or spheres in the unit cube. 
In~\citet{Conway:1998}, minimax distance designs are referred to as \emph{thinnest coverings} of the region and maximin distance designs are referred to as \emph{densest packings} with nonoverlapping balls. 
There are many results on how to place balls as $n$ goes to infinity. 
For instance, it is proved that asymptotically both the thinnest covering and the densest packing are the hexagonal lattice for $p=2$~\citep{Conway:1998}. 
A design generated by the hexagonal lattice with $n=27$ is displayed in Figure~\ref{fig:intro}(d). 
This design is almost identical to the maximin distance design with $n=27$. 
The book by~\citet{Conway:1998} gives the best known structures for $2\leq p\leq 24$. 
Designs with such structures but contained in $[0,1]^p$ are referred to as \emph{sphere packing designs} hereinafter. 
Sphere packing designs can be seen as asymptoticly minimax or maximin distance designs. 

However, similar to most maximin distance designs, the sphere packing design depicted in Figure~\ref{fig:intro}(d) is poor in projective uniformity and unsuitable for the integration purpose. 
Our solution is to rotate and translate sphere packing designs. 
We call such designs \emph{rotated sphere packing designs}. 
Rotated sphere packing designs retain same distance properties as sphere packing designs but hold greatly improved projective uniformity. 
At the very least, projections of rotated sphere packing designs are composed of distinct elements. 
Furthermore, from our construction algorithm, 
the inner points of a rotated sphere packing design have identical Voronoi cells, whose volume equals to the average volume for boundary points. 
As such, rotated sphere packing designs are also suitable for the integration purpose. 
Figure~\ref{fig:intro}(e) gives a rotated sphere packing design with $p=2$ and $n=27$. 
Note that the rotation technique has been previously applied to factorial designs and orthogonal arrays for constructing orthogonal Latin hypercube designs~\citep{Beattie:2004,Steinberg:2006,Pang:2009,Sun:2009,Sun:2010,Sun:2011}. 

The projective uniformity and integration accuracy of rotated sphere packing designs are closely related to their rotation angle. 
We detect a good angle for $p=2$, which we refer to as a ``magic'' angle. 
Rotated sphere packing designs constructed from this angle hold the quasi-Latin hypercube property~\citep{Dam:2007}. 
Specifically, when projected onto one dimension, the minimum gap distance between adjacent points is bounded above $0.289 n^{-1}$ and the maximum gap distance between adjacent points is bounded below $2.155 n^{-1}$. 
The magic angle also leads to better integration accuracy. 
It is well known that numerical integration accuracy of designs are closely related to their extreme discrepancy~\citep{Niederreiter:1992}, defined by 
\[ P(\mathbf{D}) = \sup_{\mathbf{u},\mathbf{v} \in [0,1]^p, u_1<v_1,\ldots,u_k<v_k} \left| A(\mathbf{u},\mathbf{v},\mathbf{D}) / n - \prod_{k=1}^p(v_k-u_k) \right|, \]
where $\mathbf{u}=(u_1,\ldots,u_p)$, $\mathbf{v}=(v_1,\ldots,v_p)$ and $A(\mathbf{u},\mathbf{v},\mathbf{D})$ denotes the number of points $\mathbf{x}=(x_1,\ldots,x_p)$ in $\mathbf{D}$ such that $u_k\leq x_k< v_k$ for any $k$.  
Rotated sphere packing designs with the magic angle achieve the lowest possible order of discrepancy, namely, $n^{-1}\log(n)$. 

Unfortunately, for $p\geq 3$ we do not know the optimal angle. 
Because of this problem, we propose to use the design with empirically the best projective uniformity. 
We use the criterion  
\begin{equation}
\label{eqn:criterion:MaxPro}
\psi(\mathbf{D}) = \left\{ \{n(n-1)\}^{-1}
\sum_{1\leq i<j\leq n} \frac{1}{\prod_{k=1}^p (x_{i,k}-x_{j,k})^2} \right\}^{1/p},  
\end{equation} 
which is proposed by~\citet{Roshan:2015} for generating maximum projection designs. 
Lower $\psi(\mathbf{D})$ indicates better projective uniformity.
Figure~\ref{fig:intro}(f) gives a maximum projection design with $p=2$ and $n=27$. 

Rotated sphere packing designs can be constructed easily.   
R codes to generate them for any given $p$ and $n$ are provided as supplementary material. 
Although our procedure has some optimization steps, it is much faster than those for generating maximin distance designs and maximin distance Latin hypercube designs which involve simulated annealing or other optimization techniques. 

The rest of the paper is organized as follows: 
Section~\ref{sec:lattice} gives preliminary mathematical results on lattices.  
In Section~\ref{sec:construction}, we give our construction algorithm and some theoretical results for rotated sphere packing designs. 
In Section~\ref{sec:simulation}, we compare our proposed designs with several popular classes of designs. 
Conclusions and discussion are provided in Section~\ref{sec:conclusion}. 
Proofs are given in the appendix.

\section{Preliminary results on lattices}
\label{sec:lattice}

In this section, we give some definitions and mathematical results that are necessary in constructing rotated sphere packing designs.  
Most of the results can be found in \citet{Conway:1998}.  

All best known thinnest coverings and densest packings are lattices. 
A design in $\mathbb{R}^p$ (with infinitely many points) is called a lattice if its points form a group. 
That is, if $\mathbf{u}$ and $\mathbf{v}$ are two points in the design, then any point with the form $a\mathbf{u}+b\mathbf{v}$, $a,b\in \mathbb{Z}$ is also contained in the design. 
As such, a lattice can be generated from $p$ basis vectors, $\mathbf{v_1},\ldots,\mathbf{v_p}$, with $\mathbf{v_i} = (v_{i,1},\ldots,v_{i,p})^T$, which form one of its generator matrices  
\[ \mathbf{G} = \left[ \begin{array}{cccc}
v_{1,1} & v_{1,2} & \cdots & v_{1,p} \\
v_{2,1} & v_{2,2} & \cdots & v_{2,p} \\
\vdots & \vdots & \ddots & \vdots \\
v_{p,1} & v_{p,2} & \cdots & v_{p,p} 
\end{array}\right]_{p \times p}. \]
For instance, the lattice 
\begin{equation} \label{eqn:Zm}
\{ (x_1,\ldots,x_p): x_j \in \mathbb{Z}, j=1,\ldots,p \}
\end{equation}
is called the $p$-dimensional cubic lattice $Z_p$ 
and one generator matrix for $Z_p$ is the $p\times p$ identity matrix. 

Voronoi cells from a lattice are identical, central symmetric and with volume $|\det(\mathbf{G})|$. 
The packing radius, $\rho_p$, and covering radius, $\rho_c$, of a lattice are the minimum and maximum distances from a point on the boundary of its Voronoi cell to its center, respectively. 
The density of a lattice, $\Delta$, is the volume of a ball in $\mathbb{R}^p$ with radius $\rho_p$ divided by the volume of one Voronoi cell 
and the thickness of a lattice, $\Theta$, is the volume of a ball in $\mathbb{R}^p$ with radius $\rho_c$ divided by the volume of one Voronoi cell. 
Obviously, $ 0 < \Delta \leq 1 \leq \Theta $.  
The lattice with maximum density, referred to as the densest packing, is the best solution of placing non-overlapping identical balls in $\mathbb{R}^p$ and the lattice with minimum thickness, referred to as the thinnest covering, is the best solution of placing identical balls that cover $\mathbb{R}^p$. 
We have
$ \Omega_p \rho_p^p / \Delta = |\det(\mathbf{G})| = \Omega_p \rho_c^p / \Theta ,$
where $\Omega_p$ is the volume of one unit sphere in $\mathbb{R}^p$, 
\[ \Omega_p = \frac{\pi^{p/2}}{\Gamma(p/2+1)} 
 = \begin{cases}
  \pi^{p/2}/\{(p/2)!\}, & p=2,4,6,\ldots, \\
  \pi^{(p-1)/2}2^{(p+1)/2}/(p!!), & p=1,3,5,\ldots. 
\end{cases}
\]
\citet{Conway:1998} listed best known densest packings and best known thinnest coverings up to 24 dimensions. 

If $\mathbf{G}_2 = \mathbf{G}_1 \mathbf{R}$ where $\mathbf{R}$ is an orthogonal matrix and $\det(\mathbf{R})=1$, 
the lattice generated from $\mathbf{G}_2$ can be seen as a rotation of the lattice generated from $\mathbf{G}_1$. 
We call such $\mathbf{R}$ a rotation matrix. 
When $p=2$, any rotation matrix $\mathbf{R}$ can be expressed as 
\[ \mathbf{R} = \left[ \begin{array}{cc}
\cos(\alpha) & -\sin(\alpha) \\
\sin(\alpha) & \cos(\alpha)  
\end{array}\right]  . \]
More generally, a Givens rotation $\mathbf{R}_p(i,j,\alpha)$ is the $p\times p$ identity matrix with the $(i,i)$th, $(i,j)$th, $(j,i)$th and $(j,j)$th elements being replaced by $\cos(\alpha)$, $-\sin(\alpha)$, $\sin(\alpha)$ and $\cos(\alpha)$, respectively. 
Any rotation matrix can be uniquely obtained from $p(p-1)/2$ sequential Givens rotations with $1\leq i<j\leq p$. 

For $2\leq p \leq 22$, the best known thinnest coverings are $A_p^*$, the dual of $p$-dimensional zero-sum root lattice with one possible generator matrix
\begin{eqnarray}\label{eqn:dzsrl}
\mathbf{G} &=&\frac{\sqrt{p+1}}{\sqrt{p}} \mathbf{I}_p - \frac{1}{\sqrt{p}(\sqrt{p+1}-1)} \mathbf{J}_p, 
\end{eqnarray}
where $\mathbf{I}_p$ is the $p\times p$ identity matrix and $\mathbf{J}_p$ is the $p\times p$ matrix with all elements being one. 
Using this definition, 
$ |\mathbf{v}_1| = \cdots = |\mathbf{v}_p| = 1, $
$ \rho_p = 1/2, $
$ \rho_c = \sqrt{(p+2)/12}, $
$ \det(\mathbf{G}) = -(p+1)^{(p-1)/2} p^{-p/2}, $
and $ \Theta = \Omega_p \sqrt{p+1} \left[ \{p(p+2)\}/\{12(p+1)\} \right]^{p/2}$.  
Let 
\begin{equation}\label{eqn:eta}
\boldsymbol{\eta}_j^T = \left\{ \mathbf{I}_p - \mathbf{G}_{(j)}^T ( \mathbf{G}_{(j)} \mathbf{G}_{(j)}^T )^{-1} \mathbf{G}_{(j)} \right\} \mathbf{v}_j, 
\end{equation} 
where $\mathbf{G}_{(j)}$ is the $(p-1)\times p$ matrix consisting of rows $\{\mathbf{v}_k^T:k\neq j\}$.  
Then it is not hard to show that $|\boldsymbol{\eta}_j|=(p+1)/(2p)$ for any $j$. 
The $A_2^*$ is also called the hexagonal lattice. 
Using the definition by (\ref{eqn:dzsrl}), the generator matrix is 
\begin{equation}\label{eqn:G2}
\mathbf{G}_2 = \left[ \begin{array}{ccc}
 (\sqrt{3} -1)/(2\sqrt{2}) & -(\sqrt{3} +1)/(2\sqrt{2})    \\
-(\sqrt{3} +1)/(2\sqrt{2}) &  (\sqrt{3} -1)/(2\sqrt{2})
\end{array}\right], 
\end{equation}
with 
$ \det(\mathbf{G}_2) = -\sqrt{3}/2, $
$ \rho_c = \sqrt{3}/3 $
and $ \Theta = 2\sqrt{3}\pi/9$. 

Table~\ref{tab:thickness} gives the thickness of $A_p^*$ and $Z_p$ by (\ref{eqn:Zm}) with $2\leq p\leq 10$. 
As can be seen from the table, $A_p^*$ are much more efficient than $Z_p$. 

\begin{table}
\begin{center}
\caption{Thickness of two lattices, $A_p^*$ and $Z_p$}
\begin{tabular}{cccccccccc}
\hline
\hline
$p$ & 2 & 3 & 4 & 5 & 6 & 7 & 8 & 9 & 10 \\
\hline
$A_p^*$ & 1.21 & 1.46 & 1.77 & 2.12 & 2.55 & 3.06 & 3.67 & 4.39 & 5.25 \\
$Z_p$   & 1.57 & 2.72 & 4.93 & 9.20 & 17.4 & 33.5 & 64.9 & 126.8 & 249.0 \\
\hline
\end{tabular}\label{tab:thickness}
\end{center}
\end{table}

\section{Construction and properties}
\label{sec:construction}

In this section, we provide our construction algorithm and some theoretical results behind the algorithm.

\subsection{Construction}

We now give the algorithm for generating rotated sphere packing designs. 
Justifications on the choice of parameters are given in subsequent subsections. 
The algorithm has five major steps: 
\begin{enumerate}
\item Choose a generator matrix $\mathbf{G}$ and a rotation matrix $\mathbf{R}$  
and compute 
\begin{equation}\label{eqn:l}
l = \left(n \Omega_p/\Theta\right)^{1/p} \rho_c.   
\end{equation}
\item Obtain a large design given by $\mathbf{E} = \mathbf{F} \mathbf{G} \mathbf{R}$, 
where $\mathbf{F}$ is an integer matrix sufficiently large such that 
$\{ \mathbf{f}^T\mathbf{G}\mathbf{R} : \mathbf{f} \in \mathbb{Z}^p, \mathbf{f}^T\mathbf{G}\mathbf{R} \in [-l/2-\rho_c,l/2+\rho_c]^p \}$ is a subset of rows of $\mathbf{E}$.  
One such $\mathbf{F}$ is the full factorial array with $2s+1$ levels from $-s$ to $s$, where 
\begin{equation}\label{eqn:s}
s=  \lceil (l\sqrt{p}/2+\rho_c) / \min_j(|\boldsymbol{\eta}_j|) \rceil,
\end{equation}
$\lceil z \rceil$ denotes the smallest integer no less than $z$ and $\boldsymbol{\eta}_j$ is defined by (\ref{eqn:eta}).  \item Search for a perturbation vector $\boldsymbol{\delta} =(\delta_1,\ldots,\delta_p)^T \in \text{Vor}(0)$ such that 
there are exactly $n$ points of $\mathbf{E}$ contained in the region $\otimes_{k=1}^p [-l/2-\delta_k,l/2-\delta_k]$, 
where $\text{Vor}(0)$ is the Voronoi cell of $(0,\ldots,0) \in \mathbf{E}$. 
\item Obtain the design $\mathbf{D}$ by extracting points of $ \mathbf{\tilde E}/l+1/2 $ that lie in $[0,1]^p$,  
where $\mathbf{\tilde E}$ is the matrix obtained by adding $\boldsymbol{\delta}^T$ to rows of $\mathbf{E}$. 
\item Repeat Steps~1-4 for $w$ times and select the design which minimizes $\psi(\mathbf{D})$ in (\ref{eqn:criterion:MaxPro}).  
\end{enumerate}

We recommend to use the $A_p^*$ by (\ref{eqn:dzsrl}) 
in Step~1 because they are the best known thinnest coverings. 
Designs constructed from these generator matrices are nearly minimax as $n$ goes to infinity.  
For $p\geq 3$, we recommend to use $w=100$ and generate $\mathbf{R}$ from 
\begin{equation}\label{eqn:R-Givens}
\mathbf{R} = \prod_{1\leq i<j\leq p} \mathbf{R}_p(i,j,\alpha_{i,j}),  
\end{equation}
where $\mathbf{R}_p(i,j,\alpha_{i,j})$ are Givens rotations defined in Section~\ref{sec:lattice}
and $\alpha_{i,j}$ are generated randomly and independently from the uniform distribution on $[0,2\pi)$.
From our experience, there is little gain in using $w>100$. 
When $p=2$, however, we recommend to use $w=1$ and $\mathbf{R}=\mathbf{I}_2$. 
Step~3 is the key step in getting designs with any $p$ and $n$. 

\begin{figure}
\begin{center}
\includegraphics[width=\textwidth]{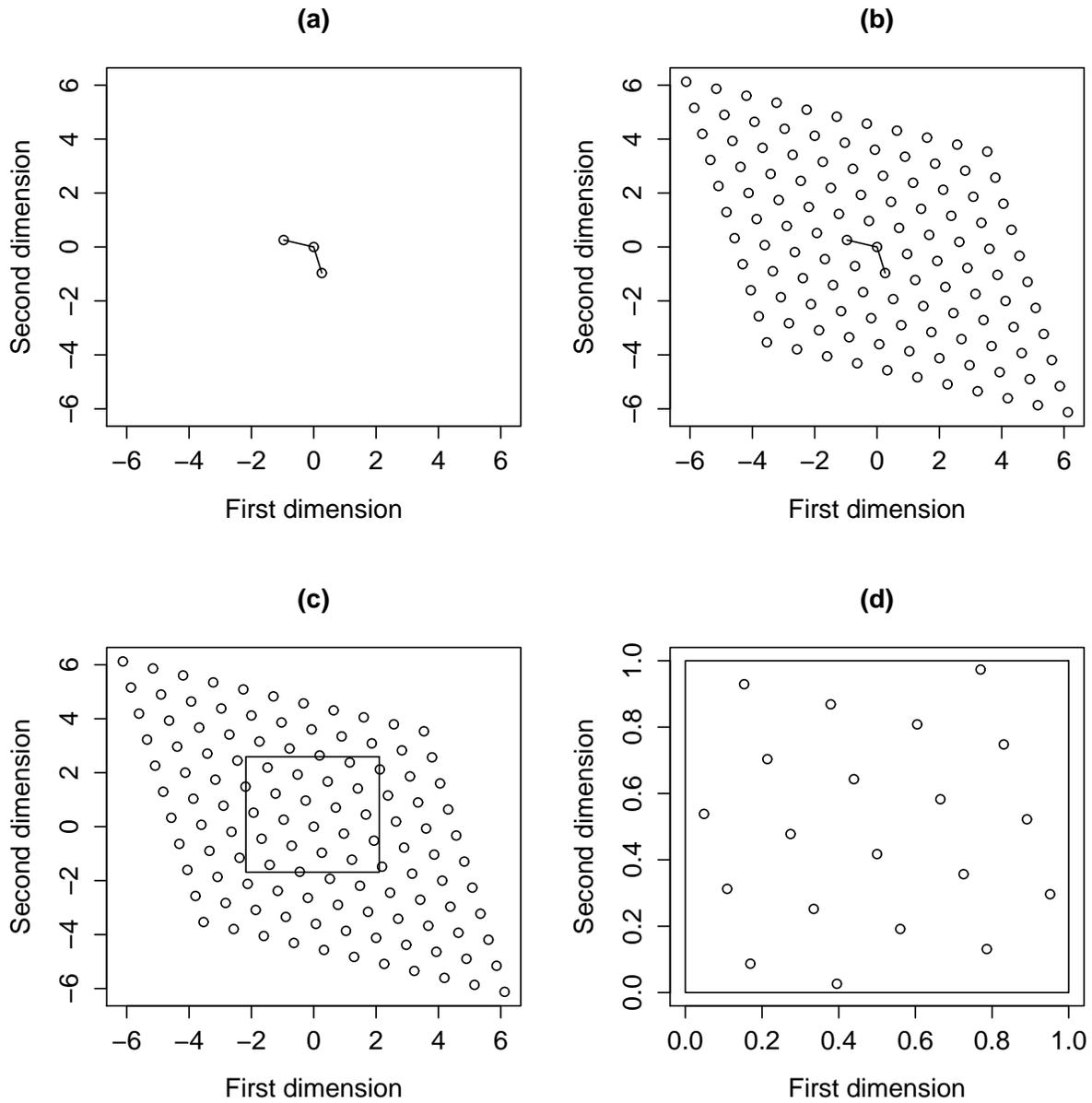}
\caption{The steps to generate a rotated sphere packing design, $p=2$, $n=20$. Plots (a), (b), (c) and (d) display designs obtained after Steps~1, 2, 3 and 4 of the algorithm, respectively. \label{fig:algo}}
\end{center}
\end{figure}

We now demonstrate the algorithm by an example with $p=2$ and $n=20$. 
In Step~1, we use $\mathbf{G}_2$ in (\ref{eqn:G2}) and $\mathbf{R}=\mathbf{I}_2$. 
Row vectors of $\mathbf{G}_2\mathbf{R}$ are $(0.259,-0.966)$ and $(-0.966,0.259)$, which are indicated by lines starting from the origin in Figure~\ref{fig:algo}(a). 
Using the properties of $A_p^*$ summarized in Section~\ref{sec:lattice}, we compute $l=4.28$ and $s=5$. 
We then generate $121\times 2$ matrices $\mathbf{F}$ and $\mathbf{E}$.  
Rows of $\mathbf{E}$ are indicated by points in Figure~\ref{fig:algo}(b).
In Step~3, we find a vector $\boldsymbol{\delta}=(0.037, -0.453)^T$ such that there are exactly 20 points of $\mathbf{E}$ contained in the region $\otimes_{k=1}^p [-l/2-\delta_k,l/2-\delta_k]$. 
Rows of $\mathbf{E}$ and the region are indicated by points and the square in Figure~\ref{fig:algo}(c), respectively. 
The searching of suitable $\boldsymbol{\delta}$ is further discussed below Theorem~\ref{thm:R}. 
Finally, we compute $\mathbf{\tilde E}$ and $\mathbf{D}$ with rows of $\mathbf{D}$ indicated by points in Figure~\ref{fig:algo}(d). 
Figure~\ref{fig:algo} illustrates the four steps. 

\subsection{General theoretical results}\label{sec:theory}

We now give some theoretical results that provide guidance to the proposed algorithm. 
The first result is concerned with $l$ in (\ref{eqn:l}). 

\begin{theorem}\label{thm:l}
Assume $l=\left(n \Omega_p/\Theta\right)^{1/p} \rho_c$. 
Then for any point whose Voronoi cell locate inside $[0,1]^p$, the volume of its Voronoi cell is $1/n$. 
\end{theorem}

Unless noted otherwise, proofs are given in the appendix.
We call the points whose Voronoi cell located inside $[0,1]^p$ as ``inner points'' and the others as ``boundary points''.
From Theorem~\ref{thm:l}, for any $n$, the boundary points and inner points on average have the same volume of Voronoi cells and thus designs generated from the proposed algorithm have no boundary problem and low discrepancy. 
The next result is concerned with the choice of $s$ in (\ref{eqn:s}). 

\begin{theorem}\label{thm:s}
Assume $\mathbf{f}=(f_1,\ldots,f_p)^T \in \mathbb{Z}^p$ and 
there is a $j$ such that $|f_j| > l\sqrt{p} / (2|\boldsymbol{\eta}_j|) + \rho_c/|\boldsymbol{\eta}_j|$, where $\boldsymbol{\eta}_j$ is defined by (\ref{eqn:eta}). 
Then for any $\boldsymbol{\delta} \in \text{Vor}(0)$, $ \mathbf{f}^T\mathbf{G}\mathbf{R}+\boldsymbol{\delta}^T \notin [-l/2,l/2]^p. $
\end{theorem}

A lattice by definition includes infinitely many points. 
Nevertheless, in light of Theorem~\ref{thm:s}, we only need to consider the points generated from $\mathbf{f}^T\mathbf{G}\mathbf{R}$ with $\mathbf{f}\in \{-s,-s+1,\ldots,s\}^p$ and $s= \lceil (l\sqrt{p}/2+\rho_c) / \min_j(|\boldsymbol{\eta}_j|) \rceil$ since other points are guaranteed to be excluded from the final design $\mathbf{D}$. 
The next result is concerned with the choice of $\mathbf{R}$. 

\begin{theorem}\label{thm:R}
Let $\mathbf{e}_j=(0,\ldots,1,\ldots,0)^T$ be the $p$-vector with the $j$-th element being one 
and let $\mathbf{S}_j = \{ \mathbf{f}^T \mathbf{G} \mathbf{R} \mathbf{e}_j : \mathbf{f} \in \{-2s,\ldots,2s\}^p, \mathbf{f} \neq 0 \}$. 
Assume $0 \notin \mathbf{S}_j$ for any $j$. 
Then the one-dimensional projections of $\mathbf{E}$ and $\mathbf{D}$ have nonidentical elements. 
Furthermore, there is at least a $\boldsymbol{\tilde\delta} \in \text{Vor}(0)$ such that there are exactly $n$ points of $\mathbf{E}$ contained in the region $\otimes_{k=1}^p [-l/2-\delta_k,l/2-\delta_k]$. 
\end{theorem}

Theorem~\ref{thm:R} is the key result in getting designs with any $p$ and $n$. 
In light of Theorem~\ref{thm:R}, we shall choose $\mathbf{R}$ so that $0 \notin \mathbf{S}_j$ for any $j$. 
Note that if $p\geq 3$ and we use the generator matrix in (\ref{eqn:dzsrl}) and $\mathbf{R}=\mathbf{I}_p$, then $0 \in \mathbf{S}_j$ for any $j$. 
Thus, when $p\geq 3$, $\mathbf{R}$ should not be the identity matrix. 
Fortunately for us, if we generate $\mathbf{R}$ randomly from (\ref{eqn:R-Givens}), then $0 \notin \mathbf{S}_j$ for any $j$ with probability one. 
These $\mathbf{R}$s are clearly not optimal but we currently have no better solution. 
Nevertheless, because of Step~5 of the construction algorithm, final designs are usually associated with good $\mathbf{R}$s. 

In light of the proof of Theorem~\ref{thm:R}, we can always convert the problem of searching a suitable $\boldsymbol{\tilde\delta}$ into a one-dimensional optimization problem. 
The search can be quick without regenerating $\mathbf{E}$ as in our R codes, but we omit the details here.

\subsection{The magic angle for $p=2$}\label{sec:p=2}

Recall that when $p=2$ we recommend to use the generator matrix $\mathbf{G}_2$ in (\ref{eqn:G2}) and $\mathbf{R}=\mathbf{I}_2$. 
This angle performs surprisingly good and thus we call it a magic angle. 
Because this angle guarantees good projective uniformity, there is no need to regenerate designs in Step~5 of the construction algorithm and therefore we recommend to use $w=1$. 
In this subsection we present some results on designs generated from this magic angle. 
Note that $\mathbf{G}_2$ can be seen as a 15 degree rotation from the hexagonal lattice presented in Figure~\ref{fig:intro}(d). 

We begin by analyzing how close two design points can be in the one-dimensional projections. 
For any $k=0,1,2,\ldots$, let 
\begin{eqnarray*}
\mathbf{y}_{2k+1}^T &=& \left( \frac{\left(\sqrt{3}+1\right)^{2k}-\left(\sqrt{3}-1\right)^{2k}}{2^{k+1}\sqrt{3}}, 
 \frac{\left(\sqrt{3}+1\right)^{2k+2}-\left(\sqrt{3}-1\right)^{2k+2}}{2^{k+2}\sqrt{3}} \right) \mathbf{G}_2 l \\
 &=&  \left( -(\sqrt{3}+1)^{2k+1}/2^{k+3/2}, (\sqrt{3}-1)^{2k+1}/2^{k+3/2} \right) /l ,\\
\mathbf{y}_{2k+2}^T &=& \left( \frac{-\left(\sqrt{3}+1\right)^{2k+1}-\left(\sqrt{3}-1\right)^{2k+1}}{2^{k+1}\sqrt{3}}, 
 \frac{-\left(\sqrt{3}+1\right)^{2k+3}-\left(\sqrt{3}-1\right)^{2k+3}}{2^{k+2}\sqrt{3}} \right) \mathbf{G}_2/l \\
 &=&  \left( (\sqrt{3}+1)^{2k+2}/2^{k+3/2}, (\sqrt{3}-1)^{2k+2}/2^{k+3/2} \right) /l ,
\end{eqnarray*}
and write $\mathbf{y}_k^T = (y_{k,1},y_{k,2})$. 

\begin{proposition}\label{prop:p=2}
Suppose $ \mathbf{x}_0^T = (x_{0,1},x_{0,2}) = \mathbf{f}_0^T \mathbf{G}_2 /l$, $\mathbf{f}_0^T=(f_{0,1},f_{0,2}) \in \mathbb{Z}^2 \setminus \{0,0\}$,  $k\in\mathbb{N}$ and $|x_{0,1}|<|y_{k,1}|$. 
Then $|x_{0,2}|>y_{k,2}$. 
\end{proposition}

For any two different design points $\mathbf{x}_1,\mathbf{x}_2 \in \mathbf{D}$, the difference $\mathbf{x}_0=\mathbf{x}_2-\mathbf{x}_1$ can be written as $ \mathbf{x}_0^T = \mathbf{f}_0^T \mathbf{G}_2 /l$ with an $\mathbf{f}_0 \in \mathbb{Z}^p$. 
Therefore, all scenarios in which $\mathbf{x}_1$ and $\mathbf{x}_2$ are close in the second dimension are characterized by the sequence of $(\mathbf{y}_1,\mathbf{y}_2,\ldots)$, which we refer to as the minimum vectors for $\mathbf{G}_2$ hereinafter. 
Obviously, we can reach similar results for the first dimension. 
Based on Proposition~\ref{prop:p=2}, clearly $0\notin \mathbf{S}_1$ and $0\notin \mathbf{S}_2$ and thus the one-dimensional projections of $\mathbf{D}$ has nonidentical elements. 
The theorem below gives more properties of $\mathbf{D}$. 

\begin{theorem} \label{thm:p=2}
Suppose $\mathbf{D}$ is generated with generator matrix $\mathbf{G_2}$ and rotation matrix $\mathbf{I}_2$ and $n\geq 2$. 
Then 
\begin{align*}
& (i)  & & P(\mathbf{D}) = O(n^{-1}\log(n)); \\
& (ii) & & \min_{\mathbf{x}_i,\mathbf{x}_j\in \mathbf{D},\mathbf{x}_i\neq \mathbf{x}_j} |x_{i,2}-x_{j,2}| \geq \frac{\sqrt{3}}{6} n^{-1}; \\
& (iii)& & \max \left( \min_{\mathbf{x}_j \in \mathbf{D}, x_{j,2}>x_{i,2}} |x_{j,2}-x_{i,2}| \right) 
 \leq \left( \frac{2\sqrt{3}}{3}+1 \right) n^{-1}, 
\end{align*}
where the maximum in (iii) is over $i$ such that $\mathbf{x}_i \in \mathbf{D}$ and $x_{i,2}\neq \max_{\mathbf{x}_j\in \mathbf{D}} (x_{j,2})$. 
\end{theorem}

From Theorem~\ref{thm:p=2}(i), rotated sphere packing designs generated by $\mathbf{G}_2$ and $\mathbf{R}=\mathbf{I}_2$ achieve the lowest possible rate on discrepancy~\citep{Niederreiter:1992}. 
Thus, they are one class of low-discrepancy points that are guaranteed to be good for numerical integration. 
Theorem~\ref{thm:p=2}(ii) and (iii) show that such designs achieve excellent uniformity on the one-dimensional projections and are quasi-Latin hypercube designs defined in  \citet{Dam:2007}.

\section{Numerical illustration}\label{sec:simulation}

In this section, we corroborate the effectiveness of rotated sphere packing designs by a simulation study. 
We compare rotated sphere packing designs with four popular classes of space-filling designs:  
\begin{description}
\item[RSPD] The rotated sphere packing designs with randomly generated $\mathbf{R}$ and $w=100$. 
\item[RSPDM] The rotated sphere packing designs with the magic angle ($p=2$) and $w=1$. 
\item[MMLH] The maximin distance Latin hypercube designs. 
Specifically, we use the ``maximinSLHD'' function from the R package ``SLHD''~\citep{R:SLHD} with default settings. 
\item[ULH] The uniform Latin hypercube designs generated from a simulated annealing algorithm~\citep{Fang:2000,Morris:1995}.  
Specifically, we use the ``discrepSA\_LHS'' function with the centered $L_2$ discrepancy from the R package ``DiceDesign''~\citep{R:DiceDesign} with default settings. 
\item[MPLH] The maximum projection Latin hypercube designs~\citep{Roshan:2015}, generated from the R package ``MaxPro''~\citep{R:MaxPro}. 
\item[Hamm] The Hammersley point set~\citep{Niederreiter:1992}. 
\end{description}
We have tried MMLHs generated from the R package ``DiceDesign'' and ULHs generated from JMP, 
but they did not perform better.  
We have also tried other quasi-Monte Carlo methods such as the Halton point set and the Sobol' sequence. 
However, we do not include their results because they perform similarly as Hamm. 

To compare the prediction accuracy of designs, we use three performance measures: 
\begin{description}
\item [MinDist] The minimum pairwise distance by (\ref{eqn:MinDist}).  
\item [IMSPE] The integrated mean squared prediction error, 
\[ \int_{[0,1]^p} E \left\{ ( \hat Y(x) - Y(x) )^2 \right\} dx, \] 
where $Y(x)$ is the realization of a Gaussian process and $\hat Y(x)$ is the predicted outcome from a Gaussian process model with correctly specified correlation function. 
\item [IMSPE-inner] The integrated mean squared prediction error over the inner region, 
\[ \int_{[0.1,0.9]^p} E \left\{ ( \hat Y(x) - Y(x) )^2 \right\} dx. \] 
\end{description}
For IMSPE and IMSPE-inner, we assume the Gaussian process has constant mean and the covariance between $\mathbf{x}=(x_1,\ldots,x_z)^T$ and $\mathbf{y}=(y_1,\ldots,y_z)^T$ is 
\[ \exp\left(-\theta \sum_k (x_k-y_k)^2\right) ,\]
where $\theta$ is 24.8, 8.6, 4.6, 2.9, 2.0, 1.5, 1.2, 1.0 and 0.85 when $p$ is $2, 3, \ldots, 10$, respectively. 
With these assumptions, we use the shortcut formula below to compute the exact value of IMSPE~\citep{Sacks:1989}: 
\[ \mbox{IMSPE} = 1 - \mbox{trace}\left\{ \left( \begin{array}
{cc}
0 & \mathbf{1_n}^T \\
\mathbf{1_n} & \mathbf{C} 
\end{array} \right)^{-1} 
\left( \begin{array}
{cc}
1 & \mathbf{b}^T \\
\mathbf{b} & \mathbf{B} 
\end{array} \right) \right\}, \]
where $\mathbf{1_n}$ is the $n$-vector of ones, $\mathbf{C}$ is the correlation matrix, $\mathbf{b}=(b_1,\ldots,b_n)^T$,  
\[ b_i = \prod_k \left[ \left\{ \Phi\left(\sqrt{2\theta}(1-x_{i,k})\right) - \Phi\left(\sqrt{2\theta}(-x_{i,k})\right) \right\} \sqrt{\pi/\theta} \right], \]
$\Phi$ is the cumulative distribution function of the standard normal distribution, 
$\mathbf{B}$ is an $n\times n$ matrix with the $(i,j)$th element $b_{i,j}$, and 
\begin{eqnarray*}
b_{i,j} &=& \prod_k \left[ \left\{ \Phi\left( 2\sqrt{\theta} (1-x_{i,k}/2-x_{j,k}/2) \right) -  \Phi\left( 2\sqrt{\theta} (-x_{i,k}/2-x_{j,k}/2) \right) \right\} \right. \\&& \quad \left. \sqrt{\pi/(2\theta)} \exp\left( -(x_{i,k}-x_{j,k})^2\theta/2 \right) \right]. 
\end{eqnarray*}

\begin{figure}
\begin{center}
\includegraphics[width=13cm]{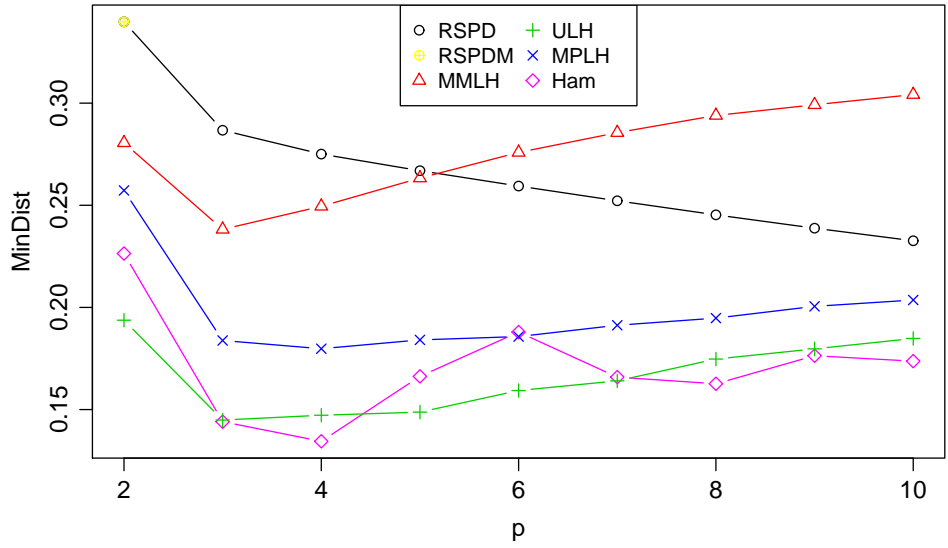}
\caption{The minimum pairwise distances for designs, the larger the better. \label{fig:MinDist}}
\end{center}
\end{figure}

\begin{figure}
\begin{center}
\includegraphics[width=13cm]{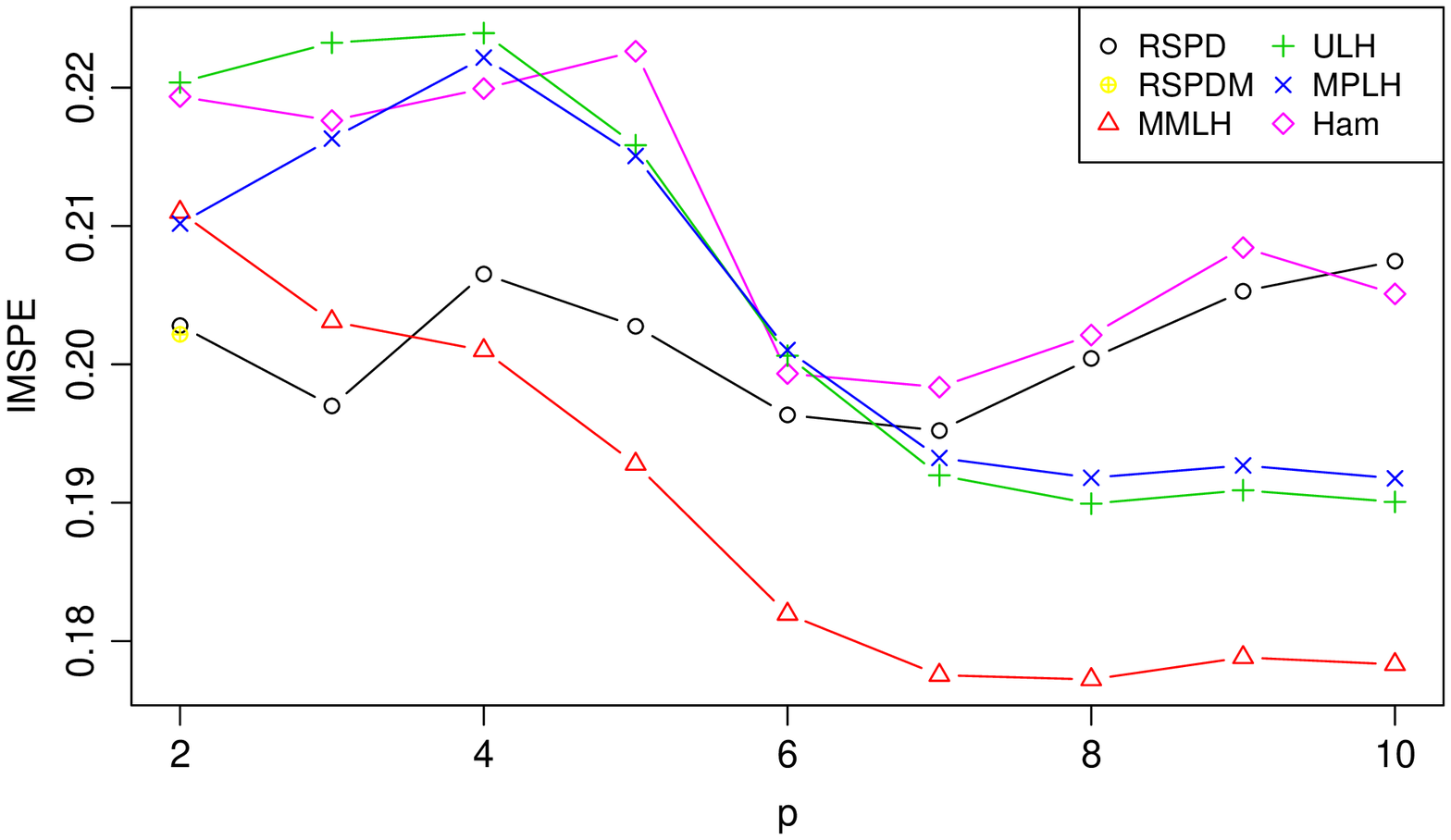}
\caption{The integrated mean squared prediction errors for designs, the smaller the better. \label{fig:IMSPE}}
\end{center}
\end{figure}

\begin{figure}
\begin{center}
\includegraphics[width=13cm]{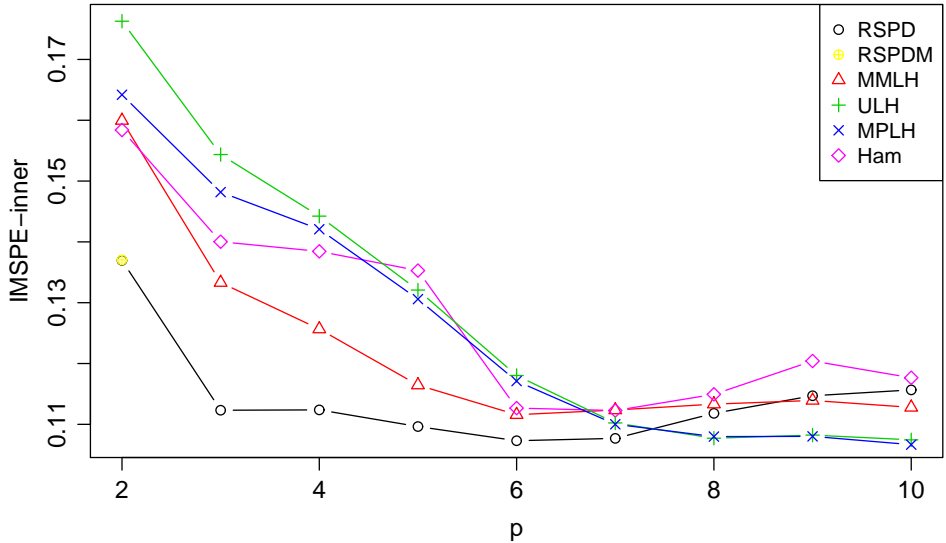}
\caption{The integrated mean squared prediction errors over $[0.1,0.9]^p$ for designs, the smaller the better. \label{fig:IMSPE9}}
\end{center}
\end{figure}

We assume $2\leq p\leq 10$ and $n=10p$. 
Figures~\ref{fig:MinDist}-\ref{fig:IMSPE9} show the simulation results. 
Generally, RSPD performs well for low dimensions and RSPDM performs as well as RSPD for $p=2$. 
MMLH is another good method which performs robustly well. 
Specifically, RSPD and RSPDM are the best methods for MinDist, IMSPE and IMSPE-inner when $p\leq 5$, $p\leq 3$ and $p\leq 7$, respectively. 
For higher $p$, RSPD is inferior to MMLH. 

We observe that RSPD performs much better for IMSPE-inner than for IMSPE. 
This implies that RSPD is more desirable for predicting the inner region than for the boundary regions. 
Some extra simulation results for scenarios with higher $n$ are shown in Figure~3 of the supplementary material. 
From the results, RSPD becomes more desirable as $n$ grows. 
For instance, RSPD is the best method for IMSPE when $p=5$ and $n\geq 400$. 

\begin{figure}
\begin{center}
\includegraphics[width=13cm]{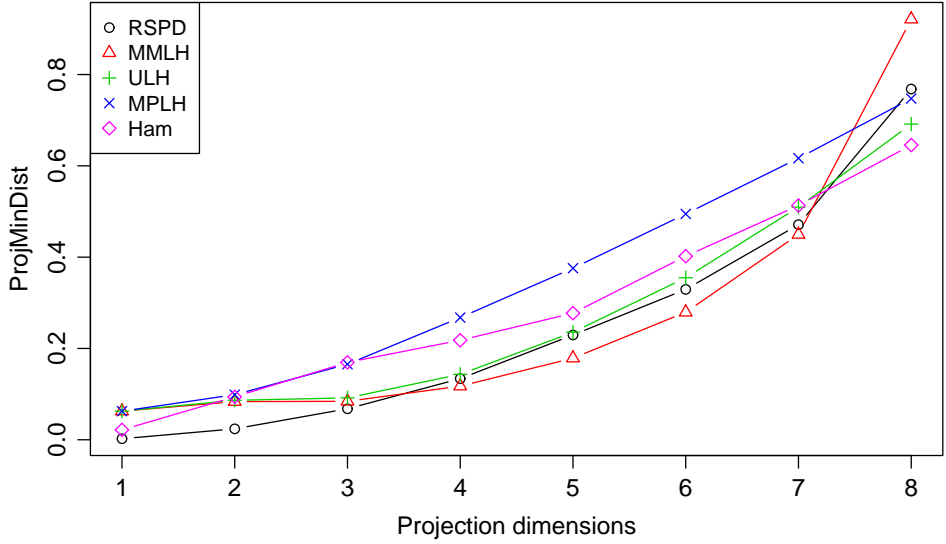}
\caption{The value of the projective minimum pairwise distance criterion by (\ref{eqn:Index}) for projections of designs, $p=8$, the larger the better. The last column gives the values for unprojected designs. \label{fig:MMD8}}
\end{center}
\end{figure}

\begin{figure}
\begin{center}
\includegraphics[width=13cm]{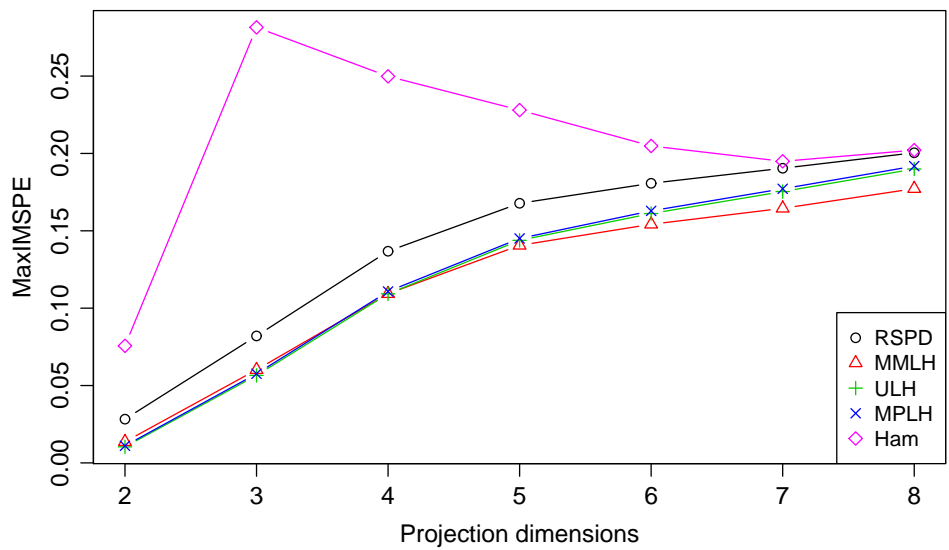}
\caption{The maximum of integrated mean squared prediction errors for projections of designs, $p=8$, the smaller the better. The last column gives the integrated mean squared prediction errors for unprojected designs.\label{fig:MIMSPE8}}
\end{center}
\end{figure}

Next, we compare the projective uniformity of these designs. 
We fix $p=8$ and $n=80$ and consider two performance measures: 
\begin{description}
\item [ProjMinDist] A criterion that measures the minimum pairwise distance of the $h$-dimensional projections of the design, introduced in~\citet{Roshan:2015} and given by 
\begin{equation} \label{eqn:Index}
\min_{u} \left\{ \frac{2}{n(n-1)} \sum_{i<j} \| \mathbf{x}_{i} - \mathbf{x}_{j} \|_u^{-2h} \right\}^{-1/(2h)}, 
\end{equation}
where the minimum is among all possible projections and $\| \mathbf{x}_{i} - \mathbf{x}_{j} \|_u$ gives the Euclidean distance between points $\mathbf{x}_i$ and $\mathbf{x}_j$ in the $u$-th projection of dimension $h$. 
\item [MaxIMSPE] The maximum of integrated mean squared prediction error for $h$-dimensional projections of the design. 
\end{description}
For MaxIMSPE, we assume the same covariance function as that for IMSPE with the same number of active dimensions.
Figures~\ref{fig:MMD8} and~\ref{fig:MIMSPE8} show the simulation results. 

For ProjMinDist, MPLH is generally the best method. 
Although being poor for $h=1,2$, RSPD has comparable performance for $h\geq 3$. 
MMLH, while being the best method for $h=p=8$, performs the worst on 4 to 7-dimensional projections. 
For MaxIMSPE, all Latin hypercube designs perform well, RSPD is slightly inferior to them and Hamm is the worst method. 
Although RSPD is not great for projective uniformity, there is a huge improvement from unrotated sphere packing designs. 

\begin{figure}
\begin{center}
\includegraphics[width=13cm]{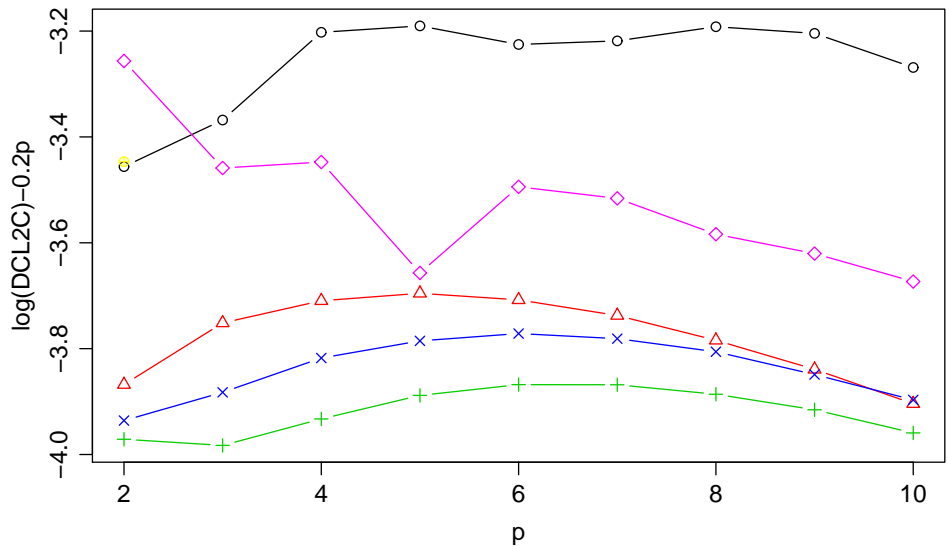}
\caption{The centered $L_2$ discrepancy for designs, the smaller the better. \label{fig:DCL2C}}
\end{center}
\end{figure}

\begin{figure}
\begin{center}
\includegraphics[width=13cm]{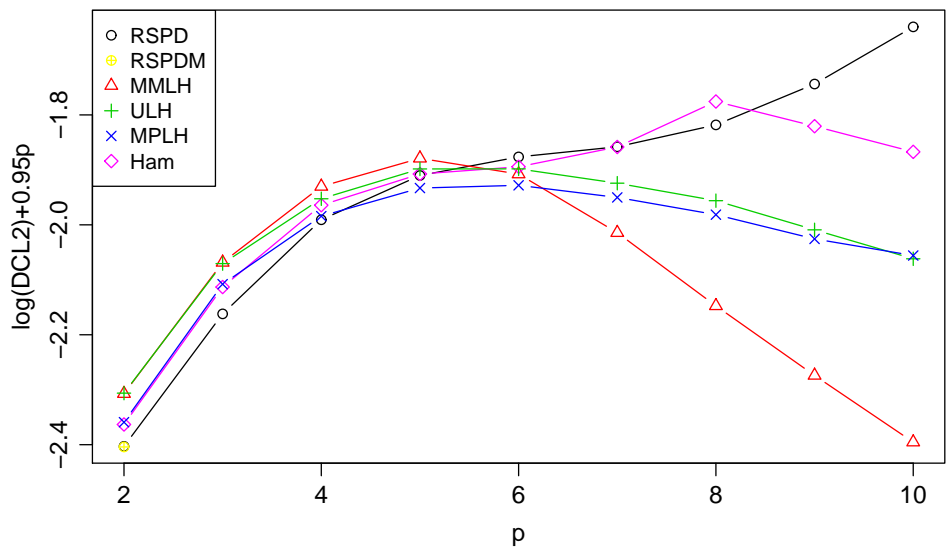}
\caption{The $L_2$ discrepancy for designs, the smaller the better. \label{fig:DCL2}}
\end{center}
\end{figure}

Then, we compare methods on their integration accuracy. 
Again, we assume $2\leq p\leq 10$ and $n=10p$. 
Figures~\ref{fig:DCL2C} and~\ref{fig:DCL2} give the simulation results for two discrepancy measures: 
\begin{description}
\item [DCL2C] The centered $L_2$ discrepancy~\citep{Niederreiter:1992}, \[ P_2^c(D) = \int_{[0,1]^p } \left| A(a(u),u,D) / n - \prod_{k=1}^p \left| u_k-a(u_k) \right| \right|,\]
where $a(u)=(a(u_1),\ldots,a(u_p))$, $a(u_k)=0$ if $u_k<1/2$ and $a(u_k)=1$ if $u_k>1/2$. 
\item [DCL2] The $L_2$ discrepancy~\citep{Niederreiter:1992}, \[ P_2(D) = \int_{[0,1]^p \otimes [0,1]^p} \left| A(u,v,D) / n - \prod_{k=1}^p(v_k-u_k) \right|. \]
\end{description}

From the results, ULH is the best method for DCL2C and MPLH is the second best. 
Although being poor for DCL2C, PSPD is good for DCL2 with low $p$. 
The difference between DCL2C and DCL2 lies in that the former considers hyperrectangles starting from one corner and the latter considers arbitrary hyperrectangles inside $[0,1]^p$. 
From the fact that RSPD performs much better for DCL2 than for DCL2C, we infer that RSPD achieves better uniformity for the inner region than for boundary regions. 
RSPD is clearly not the best choice for numerical integration of arbitrary functions, 
but it is suitable for functions who are flat on boundary regions. 
Two such examples are the continuous integrand family~\citep{Genz:1984}, 
\[ f(\mathbf{x}) = \exp\left( -\sum(5 |x_k-d_k|) \right),\]
and the Gaussian peak integrand family~\citep{Genz:1984}, 
\[ f(\mathbf{x}) = \exp\left( -\sum\{5 (x_k-d_k)^2\} \right).\]
In Figures~1 and 2 of the supplementary material, we give simulation results on estimating the mean outcome of the two families. 
From these results, RSPD and RSPDM are the best methods for the continuous integrand family when $2\leq p\leq 6$ and RSPD is the best method for the Gaussian peak integrand family when $p=5,6,9,10$. 

We also show some simulation results for scenarios with higher $n$ in Figure~4 of the supplementary material. 
From these results, RSPD, RSPDM and Hamm perform better for integration as $n$ grows. 
Some other results suggest RSPDM is the best method for DCL2C when $p=2$ and $n\geq 660$ and Hamm is the best method for DCL2C when $3\leq p\leq 10$ and $n\geq 500$. 

\begin{figure}
\begin{center}
\includegraphics[width=13cm]{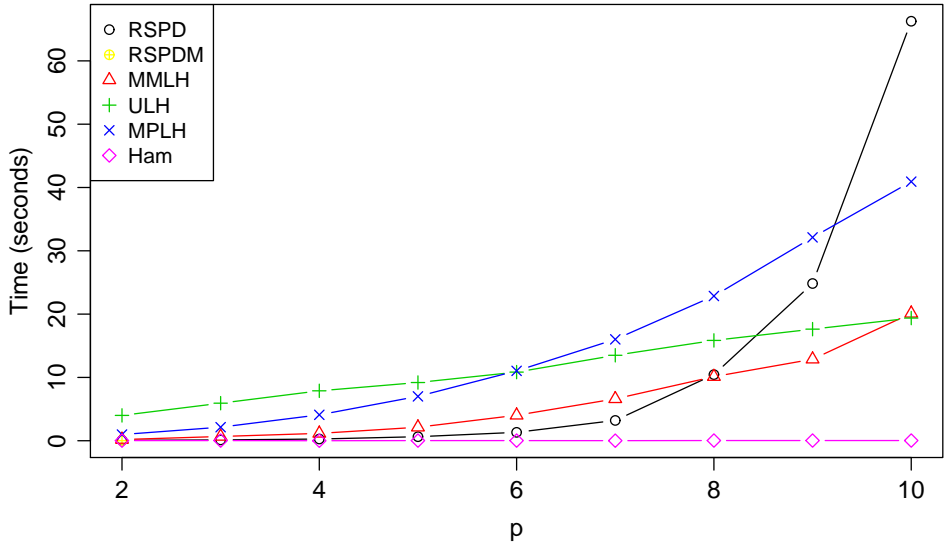}
\caption{The computation times (in seconds) for generating one design, the smaller the better. \label{fig:time}}
\end{center}
\end{figure}

Finally, we show the computation time for generating one design with $2\leq p\leq 10$ and $n=10p$ in Figure~\ref{fig:time}.  
Hamm is the fastest method and RSPD is also very fast for $2\leq p\leq 7$. 
However, it takes much longer to generate an RSPD for $p>8$. 
This is because for large $p$, the matrix $\mathbf{E}$ contains a huge number of rows which makes Step~2 of the construction algorithm slow.  
This fact again suggests that RSPD is not suitable for large $p$ problems. 
Also note that Hamm allows points to be added one-at-a-time while other methods do not. 

\section{Conclusions and further discussion}\label{sec:conclusion}

We propose a new class of space-filling designs called rotated sphere packing designs. 
Such designs are constructed by exploiting existing mathematical results of placing identical balls in a unit cube. 
By rescaling, rotating, translating and extracting the asymptotically best placement of balls, we develop an algorithm to generate rotated sphere packing designs for any number of dimensions and points. 
R codes for the algorithm are provided as supplementary material. 
The construction algorithm is very fast. 
This is an advantage over the popular maximin distance Latin hypercube designs which involve time-consuming optimization steps. 
Therefore, rotated sphere packing designs can be used as good starting designs if other optimization based algorithms are available for generating minimax distance designs. 

Unlike various kinds of Latin hypercube designs, rotated sphere packing designs possess perfect local structures.  
One consequence is that we can scale rotated sphere packing designs such that the inner points and boundary points on average have the same volume of Voronoi cells. 
Therefore, points from rotated sphere packing designs can be regarded as uniformly distributed in $[0,1]^p$.
This in turn makes such designs desirable in estimating the mean response. 
Particularly, designs generated with the magic angle are low-discrepancy points. 
Therefore, rotated sphere packing designs are useful in numerical integration~\citep{Kuo:2011}, stochastic optimization~\citep{Shapiro:2009} and uncertainty quantification~\citep{Xiu:2010}. 

Rotated sphere packing designs achieve distance properties because they are constructed based on the best known thinnest coverings. 
In other words, rotated sphere packing designs can be seen as asymptotically minimax distance designs. 
For finite $n$, they are nearly minimax for the inner region because the worst predictive distance in (\ref{eqn:MaxDist}) is perfectly controlled. 
However, as we can see from Figure~\ref{fig:intro}(e), predictive distances for the boundary (and especially the corner) regions are not controlled and can be poor. 
This fact greatly impact the prediction accuracy of rotated sphere packing designs over the whole $[0,1]^p$
region. 
We plan to investigate on ways to improve the boundary uniformity of rotated sphere packing designs in the future. 

It is discussed in \citet{Dette:2010} that designs with more points in the boundaries tend to perform better in integrated prediction accuracy because designs whose points are uniformly distributed in the design space have difficulty predicting the boundary regions. 
As a result, \citet{Dette:2010} recommended to transform a uniformly distributed design into a nonuniform design to improve the overall prediction accuracy. 
Under this framework, uniformly distributed designs with distance-based uniformity and/or predicting the inner region well may be desirable. 

Rotated sphere packing designs achieve much better projective uniformity than maximin distance designs by coupling a rotation step. 
The rotation step also makes it possible to generate designs with any number of inputs. 
While a random rotation is better than no rotation, it would be ideal that we use the best rotation angle. 
Unfortunately, we only know a good angle for the two-dimensional case which are not proved to be optimal in any sense. 
We call it a magic angle because designs generated from this angle are quasi-Latin hypercube designs. 
A future research problem is to find more magic angles for higher dimensions. 
For the case with more than two dimensions, we propose to generate multiple designs and select the one with the best projective uniformity. 
The empirically best designs are usually associated with good angles. 

Although we focus at the best known thinnest coverings in this paper, our construction method are also applicable to other lattices. 
For instance, we may use the best known densest packings listed in~\citet{Conway:1998}. 
We expect to obtain nearly maximin distance designs from these lattices. 
We plan to study such designs in a future project. 

Although we focus on the unit cube, rotated sphere packing designs are potentially useful for other design spaces. 
Certainly some modifications on the construction algorithm are needed. 
For instance, we shall extract points that lie in the design space rather than in $[0,1]^p$ in Step~4. 

Rotated sphere packing designs have explicit mathematical formulations which makes it convenient in deriving theoretical results. 
For instance, the worst predictive distance for non-boundary regions is a function of the number of dimensions and points.  
Therefore, we can calculate the sample size needed to reach a prespecified worst predictive distance. 
This provides an alternative choice to the popular rule of 10 times the number of dimensions~\citep{10d}. 

The major restriction of rotated sphere packing designs is on the number of dimensions. 
Seen from numerical results, rotated sphere packing designs are excellent for two to five dimensions. 
However, as the number of dimensions grows, they become inferior to existing methods such as maximin distance Latin hypercube designs. 
Clearly, rotated sphere packing designs are not suitable for large $p$ problems.

\section*{Appendix}

\textbf{Proof of Theorem~\ref{thm:l}}

\begin{proof}
Recall from Section~\ref{sec:lattice} that the volume of Voronoi cells for a lattice generated by $\mathbf{G}$ is 
$ \det(\mathbf{G}) = \Omega_p \rho_c^p / \Theta $. 
Therefore, the volume of Voronoi cells for a lattice generated by $\mathbf{G}/l$ is $\det(\mathbf{G})/l^p = 1/n$. 
For a point whose Voronoi cell locates inside $[0,1]^p$, none of the rotation, translation and extraction steps change the volume of its Voronoi cell and thus the volume of its Voronoi cell is the same to that of the lattice generated by $\mathbf{G}/l$. 
\end{proof}

\textbf{Proof of Theorem~\ref{thm:s}}

\begin{proof}
From (\ref{eqn:eta}), we can write 
\[ \mathbf{v}_j = \boldsymbol{\eta}_j + \sum_{k\neq j} \beta_k \mathbf{v}_k, \]
where for any $k\neq j$, $\boldsymbol{\eta}_j^T \mathbf{v}_k=0$. 
Thus, 
\[ |\mathbf{f}^T\mathbf{G}| = \left| \sum_k (f_k\mathbf{v}_k) \right| 
 = \left| f_j\boldsymbol{\eta}_j + \sum_{k \neq j} (f_j\beta_k + f_k)\mathbf{v}_k \right|
 \geq |f_j| |\boldsymbol{\eta}_j|. \]
From the assumption, 
\[ |\mathbf{f}^T\mathbf{G}\mathbf{R}+\boldsymbol{\delta}^T| \geq |\mathbf{f}^T\mathbf{G}| - |\boldsymbol{\delta}| \geq l\sqrt{p}/2 \]
and thus $\mathbf{f}^T\mathbf{G}\mathbf{R}+\boldsymbol{\delta}^T \notin [-l/2,l/2]^p$. 
\end{proof}

\textbf{Proof of Theorem~\ref{thm:R}}

\begin{proof}
Suppose the $j$-th column of $\mathbf{E}$ has two identical entries. 
Then there are $\mathbf{f}_1,\mathbf{f}_2 \in \{-s,\ldots,s\}^p$, $\mathbf{f}_1\neq \mathbf{f}_2$, such that  
$ \mathbf{f}_1^T \mathbf{G} \mathbf{R} \mathbf{e}_j = \mathbf{f}_2^T \mathbf{G} \mathbf{R} \mathbf{e}_j$. 
let $\mathbf{f}_3=\mathbf{f}_1-\mathbf{f}_2$. 
Then $\mathbf{f}_3 \in \{-2s,\ldots,2s\}^p, \mathbf{f}_3 \neq 0$ and   
$ \mathbf{f}_3^T \mathbf{G} \mathbf{R} \mathbf{e}_j =0$. 
Consequently, $0 \in \mathbf{S}_j$, which contradicts the assumption. 
Therefore the one-dimensional projections of $\mathbf{E}$ and $\mathbf{D}$ have nonidentical elements.  

Consider the $q^p$ cubic bins given by $\prod_k [\gamma_k +a_kl -l/2,\gamma_k +a_kl +l/2]$,  
where $a_k =1,\ldots,q$, $k=1,\ldots,p$ and $\boldsymbol{\gamma}=(\gamma_1,\ldots,\gamma_p)^T \in \mathbb{R}^p$ is a vector such that for any $\mathbf{f} \in \mathbb{Z}^p$, $\mathbf{f}^T\mathbf{G}\mathbf{R}$ are not located on the boundary of the bins. 
Let $g(\mathbf{z})$ denote the number of points of $\mathbf{f}=(f_1,\ldots,f_p)^T \in \mathbb{Z}^p$ such that $(\mathbf{f}^T\mathbf{G}\mathbf{R}+\mathbf{z}^T)$ is contained in the region $[-l/2,l/2]^p$. 
Then $g(\boldsymbol{\gamma}+l \mathbf{a})$ takes integer values when $\mathbf{a} \in \{1,\ldots,q\}^p$ 
and \[ (q-2\rho_c/l)^p n \leq \sum_{\mathbf{a} \in \{1,\ldots,q\}^p} g(\boldsymbol{\gamma}+l \mathbf{a}) \leq (q+2\rho_c/l)^p n. \]
Thus, for $q> (2\rho_c/l)\{ 1-(1-1/n)^{1/p} \}$, there is at least an $\mathbf{a}_1$ such that $g(\boldsymbol{\gamma}+l \mathbf{a}_1)\leq n$ 
and for $q> (2\rho_c/l)\{ (1+1/n)^{1/p}-1 \}$, there is at least an $\mathbf{a}_2$ such that $g(\boldsymbol{\gamma}+l \mathbf{a}_2)\geq n$. 
It is therefore not hard to see that there exists $\mathbf{a}_3,\mathbf{a}_4 \in \mathbb{Z}^p$ 
such that $g(\boldsymbol{\gamma}+l \mathbf{a}_3)\leq n$, $g(\boldsymbol{\gamma}+l \mathbf{a}_4)\geq n$ and 
$\mathbf{a}_3$ and $\mathbf{a}_4$ are different in only one dimension, say the $j$-th dimension. 
Write $\mathbf{a}_4 = \mathbf{a}_3 + b \mathbf{e}_j$. 
Obviously, $g(\boldsymbol{\gamma}+l(\mathbf{a}_3+z\mathbf{e}_j))$ is a step function for $z\in \mathbb{R}$ and for any $z_0$, 
\[ \left| \lim_{z\to z_0-}g(\boldsymbol{\gamma}+l(\mathbf{a}_3+z\mathbf{e}_j)) - \lim_{z\to z_0+}g(\boldsymbol{\gamma}+l(\mathbf{a}_3+z\mathbf{e}_j)) \right| \leq 1 .\] 
Thus, there exists a $b_0$ with $0\leq b_0\leq b$ such that $g(\boldsymbol{\gamma}+l(\mathbf{a}_3+b_0\mathbf{e}_j))=n$. 
Let $\boldsymbol{\delta}_0 = \boldsymbol{\gamma}+l(\mathbf{a}_3+b_0\mathbf{e}_j)$. 
Then there is a $\mathbf{z} \in \mathbb{Z}^p$ such that $\boldsymbol{\delta}_0 = \text{Vor}(\mathbf{z}^T\mathbf{G}\mathbf{R})$. 
Let $\boldsymbol{\tilde\delta} = \boldsymbol{\delta}_0 -\mathbf{z}^T\mathbf{G}\mathbf{R} \in \text{Vor}(0)$. 
Then $(\mathbf{E}+\boldsymbol{\tilde\delta})$ has exactly $n$ points located in the region $[-l/2,l/2]^p$. 
\end{proof}

\textbf{Proof of Proposition~\ref{prop:p=2}}

\begin{proof}
Firstly, it is not hard to see that for any $k \in \mathbb{N}$, 
$0<y_{2k+1,2}<y_{2k,2}<y_{2k-1,2}$, $-y_{2k+1,1}>y_{2k,1}>-y_{2k-1,1}>0$, 
$\mathbf{y}_{2k+1} = \mathbf{y}_{2k-1} - \mathbf{y}_{2k}$ and $\mathbf{y}_{2k+2}=\mathbf{y}_{2k}-2\mathbf{y}_{2k+1}$. 

Next, we shall show that $\mathbf{x}_0$ can be written as $\mathbf{x}_0=a_k\mathbf{y}_{k-1}+b_k\mathbf{y}_{k-2}$ with $a_k,b_k\in\mathbb{Z}$. 
Because $\mathbf{y}_1^T=(0,1)\mathbf{G}_2/l$ and $\mathbf{y}_2^T=(-1,-3)\mathbf{G}_2/l$, we can write $\mathbf{x}_0= (-3f_{0,1}+f_{0,2})\mathbf{y}_1 -f_{0,1}\mathbf{y}_2$.
Suppose $\mathbf{x}_0=a_{j-1}\mathbf{y}_{j-2}+b_{j-1}\mathbf{y}_{j-3}$. 
Then $\mathbf{x}_0 = b_{j-1}\mathbf{y}_{j-1}+(a_{j-1}+2b_{j-1})\mathbf{y}_{j-2}$ if $j$ is odd 
and $\mathbf{x}_0 = b_{j-1}\mathbf{y}_{j-1}+(a_{j-1}+b_{j-1})\mathbf{y}_{j-2}$ if $j$ is even.  
From induction on $j$, we have $\mathbf{x}_0=a_k\mathbf{y}_{k-1}+b_k\mathbf{y}_{k-2}$ with some $a_k,b_k\in\mathbb{Z}$. 
When $k>2$, by checking all possible choices of $(a_k,b_k)$, we can see that $|x_{0,2}|>y_{k,2}$.
The cases for $k=1$ and $k=2$ are trivial. 
\end{proof}

\textbf{Proof of Theorem~\ref{thm:p=2}}

Proof of Theorem~\ref{thm:p=2} is provided in the supplementary material.

\bigskip
\begin{center}
{\large\bf SUPPLEMENTARY MATERIAL}
\end{center}

\begin{description}

\item[R-codes to generate rotated sphere packing designs:] A function written in R to generate rotated sphere packing designs. 

\item[Figures for numerical comparison:] Further simulation results on the numerical comparison of methods. 

\item[Proof of Theorem~\ref{thm:p=2}:] Proof of Theorem~\ref{thm:p=2}.

\end{description}

\bibliographystyle{Chicago}
\bibliography{RSPD}

\begin{thebibliography}{}

\bibitem[\protect\citeauthoryear{Ba}{Ba}{2015}]{R:SLHD}
Ba, S. (2015).
\newblock {\em SLHD: Maximin-Distance (Sliced) {L}atin Hypercube Designs}.
\newblock R package version 2.1-1.

\bibitem[\protect\citeauthoryear{Ba and Joseph}{Ba and Joseph}{2015}]{R:MaxPro}
Ba, S. and V.~R. Joseph (2015).
\newblock {\em MaxPro: Maximum Projection Designs}.
\newblock R package version 3.1-2.

\bibitem[\protect\citeauthoryear{Beattie and Lin}{Beattie and
  Lin}{2004}]{Beattie:2004}
Beattie, S.~D. and D.~K.~J. Lin (2004).
\newblock Rotated factorial designs for computer experiments.
\newblock {\em Journal of the Chinese Statistical Association\/}~{\em 42\/}(4),
  431--450.

\bibitem[\protect\citeauthoryear{Conway and Sloane}{Conway and
  Sloane}{1998}]{Conway:1998}
Conway, J.~H. and N.~J.~A. Sloane (1998).
\newblock {\em Sphere Packings, Lattices and Groups}.
\newblock New York: Springer.

\bibitem[\protect\citeauthoryear{Dette and Pepelyshev}{Dette and
  Pepelyshev}{2010}]{Dette:2010}
Dette, H. and A.~Pepelyshev (2010).
\newblock Generalized {L}atin hypercube design for computer experiments.
\newblock {\em Technometrics\/}~{\em 52\/}(4), 421--429.

\bibitem[\protect\citeauthoryear{Fang, Lin, Winker, and Zhang}{Fang
  et~al.}{2000}]{Fang:2000}
Fang, K.~T., D.~K.~J. Lin, P.~Winker, and Y.~Zhang (2000).
\newblock Uniform design: Theory and application.
\newblock {\em Technometrics\/}~{\em 42\/}(3), 237--248.

\bibitem[\protect\citeauthoryear{Franco, Dupuy, Roustant, Damblin, and
  Iooss.}{Franco et~al.}{2014}]{R:DiceDesign}
Franco, J., D.~Dupuy, O.~Roustant, G.~Damblin, and B.~Iooss. (2014).
\newblock {\em DiceDesign: Designs of Computer Experiments}.
\newblock R package version 1.6.

\bibitem[\protect\citeauthoryear{Genz}{Genz}{1984}]{Genz:1984}
Genz, A. (1984).
\newblock Testing multidimensional integration routines.
\newblock In {\em Proc. of international conference on Tools, methods and
  languages for scientific and engineering computation}, pp.\  81--94.

\bibitem[\protect\citeauthoryear{Grosso, Jamali, and Locatelli}{Grosso
  et~al.}{2009}]{Grosso:2009}
Grosso, A., A.~Jamali, and M.~Locatelli (2009).
\newblock Finding maximin {L}atin hypercube designs by iterated local search
  heuristics.
\newblock {\em European Journal of Operational Research\/}~{\em 197\/}(2),
  541--547.

\bibitem[\protect\citeauthoryear{Jin, Chen, and Sudjianto}{Jin
  et~al.}{2005}]{Jin:2005}
Jin, R.~C., W.~Chen, and A.~Sudjianto (2005).
\newblock An efficient algorithm for constructing optimal design of computer
  experiments.
\newblock {\em Journal of Statistical Planning and Inference\/}~{\em 134\/}(1),
  268--287.

\bibitem[\protect\citeauthoryear{John, Johnson, Moore, and Ylvisaker}{John
  et~al.}{1995}]{John:1995}
John, P. W.~M., M.~E. Johnson, L.~M. Moore, and D.~Ylvisaker (1995).
\newblock Minimax distance designs in two-level factorial experiments.
\newblock {\em Journal of Statistical Planning and Inference\/}~{\em 44\/}(2),
  249--263.

\bibitem[\protect\citeauthoryear{Johnson, Moore, and Ylvisaker}{Johnson
  et~al.}{1990}]{Johnson:1990}
Johnson, M.~E., L.~M. Moore, and D.~Ylvisaker (1990).
\newblock Minimax and maximin distance designs.
\newblock {\em Journal of Statistical Planning and Inference\/}~{\em 26\/}(2),
  131--148.

\bibitem[\protect\citeauthoryear{Joseph, Dasgupta, Tuo, and Wu}{Joseph
  et~al.}{2015}]{Joseph:2015}
Joseph, V.~R., T.~Dasgupta, R.~Tuo, and C.~F.~J. Wu (2015).
\newblock Sequential exploration of complex surfaces using minimum energy
  designs.
\newblock {\em Technometrics\/}~{\em 57\/}(1), 64--74.

\bibitem[\protect\citeauthoryear{Joseph, Gul, and Ba}{Joseph
  et~al.}{2015}]{Roshan:2015}
Joseph, V.~R., E.~Gul, and S.~Ba (2015).
\newblock Maximum projection designs for computer experiments.
\newblock {\em Biometrika\/}~{\em 102\/}(2), 371--380.

\bibitem[\protect\citeauthoryear{Kuo, Schwab, and Sloan}{Kuo
  et~al.}{2011}]{Kuo:2011}
Kuo, F.~Y., C.~Schwab, and I.~H. Sloan (2011).
\newblock Quasi-{M}onte {C}arlo method for high-dimensional integration: the
  standard (weighted {H}ilbert space) setting and beyond.
\newblock {\em Anziam Journal\/}~{\em 53\/}(1), 1--37.

\bibitem[\protect\citeauthoryear{Liefvendahl and Stocki}{Liefvendahl and
  Stocki}{2006}]{Liefvendahl:2006}
Liefvendahl, M. and R.~Stocki (2006).
\newblock A study on algorithms for optimization of {L}atin hypercubes.
\newblock {\em Journal of Statistical Planning and Inference\/}~{\em 136\/}(9),
  3231--3247.

\bibitem[\protect\citeauthoryear{Loeppky, Sacks, and Welch}{Loeppky
  et~al.}{2009}]{10d}
Loeppky, J.~L., J.~Sacks, and W.~J. Welch (2009).
\newblock Choosing the sample size of a computer experiment: A practical guide.
\newblock {\em Technometrics\/}~{\em 51\/}(4), 366--376.

\bibitem[\protect\citeauthoryear{Mak and Joseph}{Mak and
  Joseph}{2016}]{Mak:2016}
Mak, S. and V.~R. Joseph (2016).
\newblock Minimax designs using clustering.
\newblock {\em Journal of Computational and Graphical Statistics\/}.
\newblock under review, arXiv:1602.03938v2.

\bibitem[\protect\citeauthoryear{Morris and Mitchell}{Morris and
  Mitchell}{1995}]{Morris:1995}
Morris, M.~D. and T.~J. Mitchell (1995).
\newblock Exploratory designs for computational experiments.
\newblock {\em Journal of Statistical Planning and Inference\/}~{\em 43\/}(3),
  381--402.

\bibitem[\protect\citeauthoryear{Niederreiter}{Niederreiter}{1992}]{Niederreiter:1992}
Niederreiter, H. (1992).
\newblock {\em Random Number Generation and Quasi-Monte Carlo Methods}.
\newblock SIAM: Philadelphia.

\bibitem[\protect\citeauthoryear{Pang, Liu, and Lin}{Pang
  et~al.}{2009}]{Pang:2009}
Pang, F., M.~Q. Liu, and D.~K.~J. Lin (2009).
\newblock A construction method for orthogonal {L}atin hypercube designs with
  prime power levels.
\newblock {\em Statistica Sinica\/}~{\em 19\/}(4), 1721--1728.

\bibitem[\protect\citeauthoryear{Sacks, Welch, Mitchell, and Wynn}{Sacks
  et~al.}{1989}]{Sacks:1989}
Sacks, J., W.~J. Welch, T.~J. Mitchell, and H.~P. Wynn (1989).
\newblock Design and analysis of computer experiments.
\newblock {\em Statistical Science\/}~{\em 4\/}(4), 409--423.

\bibitem[\protect\citeauthoryear{Santner, Williams, and Notz}{Santner
  et~al.}{2003}]{Santner:2003}
Santner, T.~J., B.~J. Williams, and W.~I. Notz (2003).
\newblock {\em The Design and Analysis of Computer Experiments}.
\newblock New York: Springer.

\bibitem[\protect\citeauthoryear{Shapiro, Dentcheva, and Ruszczynski}{Shapiro
  et~al.}{2009}]{Shapiro:2009}
Shapiro, A., D.~Dentcheva, and A.~Ruszczynski (2009).
\newblock {\em Lectures on Stochastic Programming: Modeling and Theory}.
\newblock Philadelphia: SIAM-Society for Industrial and Applied Mathematics.

\bibitem[\protect\citeauthoryear{Shewry and Wynn}{Shewry and
  Wynn}{1987}]{Shewry:1987}
Shewry, M.~C. and H.~P. Wynn (1987).
\newblock Maximum entropy sampling.
\newblock {\em Journal of Applied Statistics\/}~{\em 14\/}(2), 165--170.

\bibitem[\protect\citeauthoryear{Steinberg and Lin}{Steinberg and
  Lin}{2006}]{Steinberg:2006}
Steinberg, D.~M. and D.~K.~J. Lin (2006).
\newblock A construction method for orthogonal {L}atin hypercube designs.
\newblock {\em Biometrika\/}~{\em 93\/}(2), 279--288.

\bibitem[\protect\citeauthoryear{Sun, Liu, and Lin}{Sun
  et~al.}{2009}]{Sun:2009}
Sun, F.~S., M.~Q. Liu, and D.~K.~J. Lin (2009).
\newblock Construction of orthogonal {L}atin hypercube designs.
\newblock {\em Biometrika\/}~{\em 96\/}(4), 971--974.

\bibitem[\protect\citeauthoryear{Sun, Liu, and Lin}{Sun
  et~al.}{2010}]{Sun:2010}
Sun, F.~S., M.~Q. Liu, and D.~K.~J. Lin (2010).
\newblock Construction of orthogonal {L}atin hypercube designs with flexible
  run sizes.
\newblock {\em Journal of Statistical Planning and Inference\/}~{\em
  140\/}(11), 3236--3242.

\bibitem[\protect\citeauthoryear{Sun, Pang, and Liu}{Sun
  et~al.}{2011}]{Sun:2011}
Sun, F.~S., F.~Pang, and M.~Q. Liu (2011).
\newblock Construction of column-orthogonal designs for computer experiments.
\newblock {\em Science China Mathematics\/}~{\em 54\/}(12), 2683--2692.

\bibitem[\protect\citeauthoryear{Tan}{Tan}{2013}]{Tan:2013}
Tan, M. H.~Y. (2013).
\newblock Minimax designs for finite design regions.
\newblock {\em Technometrics\/}~{\em 55\/}(3), 346--358.

\bibitem[\protect\citeauthoryear{van Dam}{van Dam}{2008}]{Dam:2008}
van Dam, E.~R. (2008).
\newblock Two-dimensional minimax {L}atin hypercube designs.
\newblock {\em Discrete Applied Mathematics\/}~{\em 156\/}(18), 3483--3493.

\bibitem[\protect\citeauthoryear{van Dam, Husslage, den Hertog, and
  Melissen}{van Dam et~al.}{2007}]{Dam:2007}
van Dam, E.~R., B.~Husslage, D.~den Hertog, and H.~Melissen (2007).
\newblock Maximin {L}atin hypercube designs in two dimensions.
\newblock {\em Operations Research\/}~{\em 55\/}(1), 158--169.

\bibitem[\protect\citeauthoryear{Xiu}{Xiu}{2010}]{Xiu:2010}
Xiu, D. (2010).
\newblock {\em Numerical Methods for Stocahstic Computations: A Spectral Method
  Approach}.
\newblock New Jersey: Princeton University Press.

\end{thebibliography}

\end{document}